# Visible, Near-, and Mid-infrared Computational Spectrometer Enabled by Single-Spinning Film Encoder

Junren Wen[a,b,d,#], Weiming Shi[a,d,#], Cheng Gao[a,b,d], Yujie Liu[c], Shuaibo Feng[a,d], Yu Shao[a,b,d], Haiqi Gao[a,b,d], Yuchuan Shao[a,b], Yueguang Zhang[c], Weidong Shen[c,*], and Chenying Yang[a,c,*]

[a] Hangzhou Institute for Advanced Study, University of Chinese Academy of Sciences, Hangzhou, Zhejiang, 310024, China

[b] Shanghai Institute of Optics and Fine Mechanics, Chinese Academy of Sciences, Shanghai, 201800, China

[c] State key laboratory of Modern Optical Instrumentation, Department of Optical Engineering, Zhejiang University, Hangzhou, Zhejiang, 310027, China

[d] Center of Materials Science and Optoelectronics Engineering, University of Chinese Academy of Sciences, Beijing, 100049, China

\# These authors contributed equally to this work.
\* E-mail: adongszju@hotmail.com (Weidong Shen); ycheny@zju.edu.cn (Chenying Yang)
\* Telephone: +86 (571) 8795 1190 (Weidong Shen); +86 (571) 8608 7323 (Chenying Yang)

**Abstract**

Computational spectrometers are pivotal in enabling low-cost, in-situ and rapid spectral analysis, with potential applications in chemistry, biology, and environmental science. However, filter-based spectral encoding approaches typically use filter arrays, complicating the manufacturing process and hindering device consistency. By capitalizing on the polarization separation effect under oblique incidence (PSEOI), we pioneer the use of a single filter for highly efficient spectral encoding, and propose a novel computational spectrometer spanning visible to mid-infrared wavelengths by combining the Single-Spinning Film Encoder (SSFE) with deep learning-based reconstruction algorithm. The particle swarm optimization (PSO) method is employed to optimize the film configuration of SSFE, achieving low-correlation and high-complexity spectral responses under different polarizations and spinning angles, thereby enhancing both spectral resolution and accuracy of reconstruction across diverse spectral ranges. Spectral resolutions up to 0.5 nm, 2 nm, 10 nm can be realized for single-peak narrowband spectra, and 3 nm, 6 nm, 20 nm for dual-peak narrowband spectra, over the visible, near-, and mid-infrared wavelength ranges, respectively. Moreover, the proposed spectrometer demonstrates an overall 81.38% precision for the classification of 220 chemical compounds, confirming its robustness and precision in practical scenarios, along with the capability for compact, cost-effective spectroscopic solutions.

**Key words**: computational spectroscopy, deep learning, chemical compound classification, single-spinning film encoder

## 1. Introduction

Spectrometers are essential instruments for analyzing the spectral composition of light, providing critical information in fields of chemistry[1], biology[2], environmental science[3], and remote sensing[4]. In traditional designs, broadband light is divided using dispersive optical components, e.g. gratings, prisms, and optical filters, before being captured by a photodetector. These dispersive optics are particularly effective for the visible and near-infrared wavelength range, providing ultra-high spectral resolution. In contrast, the Fourier Transform Infrared (FTIR) spectrometers are commonly used for mid- and far-infrared spectral analysis. FTIR spectrometers utilize an interferometer to modulate the light and produce an interference pattern, which is then mathematically transformed into a spectrum. This method allows for the simultaneous measurement of all wavelengths, offering high-sensitivity and rapid data acquisition. However, both traditional dispersive spectrometers and FTIR systems are often hindered by the benchtop size and high production costs, limiting the applications where low-cost, miniaturization and in situ analysis are prioritized over high spectral performances.

Over the past decade, computational spectrometers have gradually emerged as a promising alternative. Recent advancements in micro- and nano-fabrication, coupled with the development of compressive sensing and artificial intelligence algorithms[5–9], propelled the evolution of the low-cost and efficient device.

For the aspect of spectral encoding, the strategies can be categorized into two main approaches, one based on filters with broad responses atop detectors and the other on detector-only configurations. For filter-based encoding strategy, optical elements capable of exhibiting diverse spectral responses are potential candidates for spectral encoders. Notable examples include quantum dots[10–13], thin films[14–18], photonic crystal slabs[19,20], and metasurfaces[21–24]. Filters are commonly configured as arrays, facilitating the rapid spectral encoding in the single-shot manner, which is beneficial for real-time applications. Detector-only configurations rely on the inherent spectral sensitivity of the detectors themselves, which can be achieved through material engineering and the integration of nanostructures. These configurations can be divided into schemes employing single-dot detectors[25–31] and detector arrays[32,33]. In the case of single-dot detectors, distinct spectral responses are typically generated by applying varying bias voltages, and such method indicates the ultimate miniaturization of computational spectrometers. However, it also has significant drawbacks. Typically, a given material responds to a limited spectral band, which constrains the applicability across broader wavelengths. The responsivity of the detector may be unstable, particularly under fluctuating temperature and humidity conditions, which limits the long-term reliability of the spectrometer. In the case of detector array-based spectral encoding, reported works are limited due to the high fabrication complexity and relatively low yield rates.

Afterwards, the unknown spectrum is recovered using the reconstruction network from the compressed measurements. Commonly, the number of spectral encoders is fewer than the discrete wavelength channels, which indicates an underdetermined problem. To address the resulting ill-posed equations, compressive sensing-based reconstruction algorithms are initially employed[10,32,34,35]. Such iterative method benefits from advanced anti-noise capabilities. However, the reconstruction speed is relatively slow, particularly when applied to spectral imaging applications. Additionally, deep learning, characterized by its data-driven approach and high-speed processing, is increasingly applied for spectral reconstruction[11,16,22,29,36–40]. It leverages the substantial computational power of graphics processing units (GPUs) to efficiently reconstruct complex spectral information, rendering it highly suitable for real-time applications.

In comparison, the filter-based computational spectrometer is more stable and technically mature, holding greater commercial potential. However, it often necessitates micro- and nano-fabrication solutions for array construction, resulting in increased manufacturing costs and more complex fabrication processes. Additionally, the spectral responses between different devices may vary due to the muti-step manufacturing approaches. Furthermore, spectral encoders that are applicable across multiple wavelength bands have rarely been reported. For instance, specific quantum dots, e.g. CdS and CdSe, struggle to exhibit strong absorption over a ultra-broad bandwidth[10]. Addressing these issues, we leverage the inherent characteristics of an individual filter to revolutionize ultra-broadband spectral encoding. In this work, we harness the polarization separation effect under oblique incidence (PSEOI) and employ the single-spinning film encoder (SSFE) with broadband spectral responses spanning the visible, near-, and mid-infrared wavelength ranges for highly efficient spectral encoding. The thickness of each layer is predesigned using particle swarm optimization (PSO) method to achieve specific spectral response requirements, e.g. low spectral correlation and high spectral complexity. SSFE is fabricated using a one-step electron beam evaporation (EBE) process, enabling large-area fabrication and repeatable manufacturing. Utilizing the deep learning-based reconstruction algorithm, spectral reconstruction can be performed at higher speed compared to traditional iterative methods. Exceptional spectral resolution is achieved across the visible, near-infrared, and mid-infrared wavelength bands. Additionally, the overall effectiveness for classifying chemical compounds within the mid-infrared wavelength range has been verified, paving the way for advanced analytical techniques in various scientific and industrial fields.

## 2. Results

### 2.1 Framework of the computational spectrometer with single-spinning film encoder and deep learning

The schematic diagram of the proposed computational spectrometer is shown in Fig. 1(a), with the polarizer, SSFE, and detector arranged sequentially. In the optical setup in Fig. 1(a), the absolute transmittance of the sample can be reconstructed, regardless of the spectrum of the illumination. On the other hand, for spectral resolution tests, the broadband light source is used in conjunction with a monochromator to generate narrowband spectra with varying center wavelengths and linewidths, as shown in Supplement 1, Fig. S1, where the sample is omitted. To achieve single-filter-based spectral encoding, we take advantage of PSEOI by utilizing SSFE and a polarizer. The polarizer facilitates the generation of two separate spectral responses, $R_P$ and $R_S$, by dividing the light into p- and s-polarizations, thereby doubling the number of spectral responses $R$. High and low refractive index materials ($TiO_2$ and $SiO_2$) are deposited alternatively on a visible to mid-infrared transparent sapphire substrate to form the 10-layer compact film stack, as shown in Fig. 1(b). By progressively spinning SSFE from 0° to a maximum value of 70°, as depicted in Fig. 1(c), a dynamic variation in spectral responses is observed. The gradual rotation alters the light filtering properties of SSFE, resulting in angle-dependent spectral encoding.

Workflow of the spectral encoding and reconstruction procedure is shown in Fig. 1(d). For spectral encoding, the intensity vector collected by the detector $\left(I_{P,0°}, ..., I_{P,70°}, I_{S,0°}, ..., I_{S,70°}\right)^T$ of an unknown spectrum $S(\lambda)$ propagating through SSFE can be expressed as

$$I_i = \int_{\lambda_{\min}}^{\lambda_{\max}} R_i(\lambda) \cdot S(\lambda) d\lambda, i = (P, 0°), ..., (P, 70°), (S, 0°), ..., (S, 70°) \tag{1}$$

By discretizing the spectral dimensions, Equation (1) can be rewritten as

$$\mathrm{I} = \mathrm{R} \cdot \mathrm{S} \tag{2}$$

where R is the precalibrated spectral response matrix. Here, the number of rows in R corresponds to distinct rotation angle combinations, while the number of columns corresponds to the discrete spectral channels.

Subsequently, deep learning is employed for efficient spectral reconstruction. The entire neural network consists of a fixed encoding network and a trainable reconstruction network, as shown in Supplement 1, Fig. S2. During training the entire neural network, spectral encoding by SSFE is substituted by the fixed encoding network, structured as an unbiased fully connected layer. The fixed weight matrix represents the precalibrated spectral response matrix R. By comparing the reconstructed spectra $\mathrm{S}'$ with the reference spectra S in the training/validating dataset, the end-to-end neural network is optimized by solving

$$(\hat{W}, \hat{\theta}) = \arg\min_{W,\theta} \|\mathrm{S}' - \mathrm{S}\|_2^2 = \arg\min_{W,\theta} \|D(\mathrm{S} \cdot \mathrm{R}) - \mathrm{S}\|_2^2 \tag{3}$$

where $W, \theta$ are the parameters (weights and biases) of the trainable reconstruction network $D$. For spectral reconstruction, the entire neural network is truncated, and $I_i$ serves as the input of the trainable neural network.

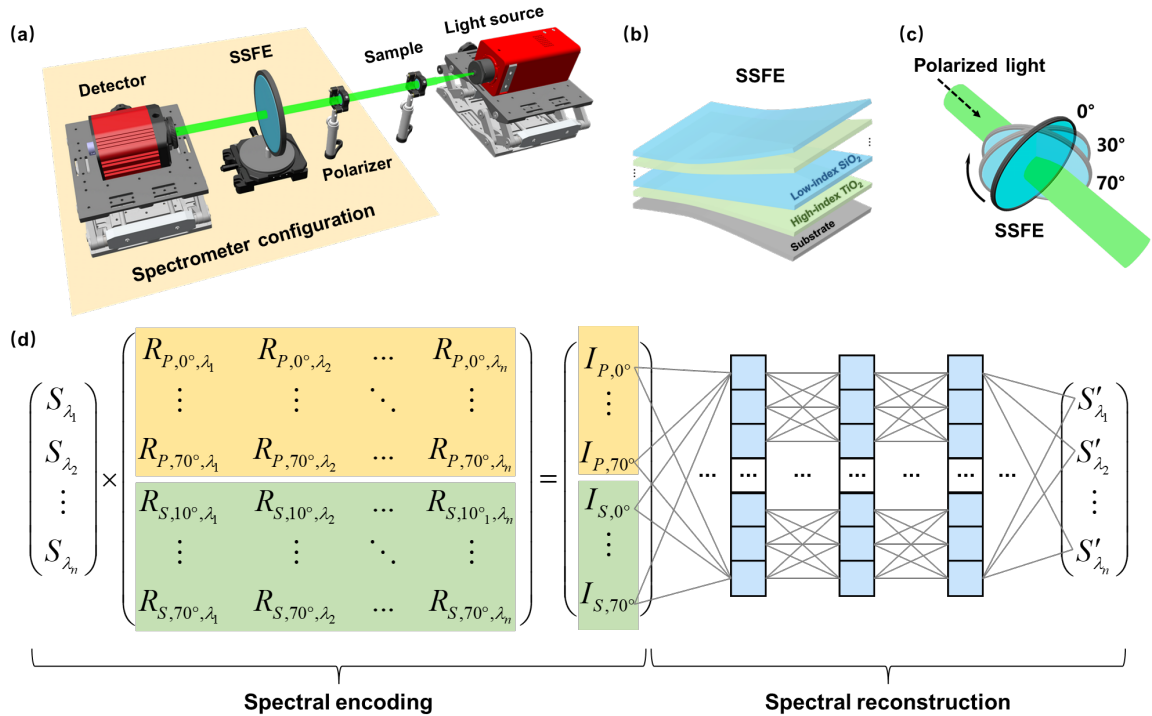

**Fig. 1.** Framework of the computational spectrometer with SSFE and deep learning. (a) The optical setup of the system, with a broadband light source, sample, and the SSFE-based spectrometer, including a polarizer, SSFE, and detector arranged sequentially. (b) Structure of the 10-layer SSFE composed of alternating high-index TiO₂ and low-index SiO₂ on a sapphire substrate. (c) Diagram illustrating the rotation of SSFE, with a rotational angle ranging from 0° to a maximum value of 70°. (d) Workflow of the spectral encoding and reconstruction procedure. Herein, spectral encoding is accomplished by SSFE and the spectral response matrix R is composed of the transmittance of SSFE under different polarizations and spinning angles. Spectral reconstruction is achieved by the deep learning-based reconstruction algorithm.

**2.2 Design of the single-spinning film encoder**

Reconstructing the unknown spectrum involves solving the underdetermined Equation (2), thereby necessitating the application of compressive sensing theory in general. Accordingly, the sensing matrix R needs to exhibit a certain level of randomness, i.e. the low correlation of the columns, to ensure stable reconstruction in principle. Since deep learning is employed for reconstruction, leveraging the advantages in handling complex spectra and accelerating reconstruction speed, it is not necessary to strictly ensure the sparse representation of the spectra under test compared with compressive sensing-based algorithms. Based on previous related works, we identify and extract two key factors for designing the SSFE: (1) low correlation coefficient of the spectral responses under different polarizations and spinning angles, and (2) high spectral complexity, characterized by the number of peaks/valleys and the corresponding amplitudes in the broadband spectral responses

    We employ the particle swarm optimization (PSO) method to design the structure of SSFE, capitalizing on the efficiency in convergence and simplicity in implementation[41,42]. Moreover, the PSO method excels in quickly reaching global optimal solutions without getting trapped in local minima, making it ideal for straightforward optimization tasks. The layer number is predetermined to be 10, with a predefined upper limit for the thickness of each individual layer. The rotation angles are set at 0°, 10°, 20°, 25°, 30°, 35°, 40°, 55°, 50°, 55°, 60°, 65°, and 70° for each polarization. Since $R_p$ and $R_s$ are identical at 0°, there are a total of 25 distinct spectral responses. During the optimization process, the 25 distinct spectral responses, corresponding to the specific layer thicknesses, are determined using the transfer matrix method[43]. These responses are then employed to calculate the Figure of Merit (FoM). Taken as an example, the FoM for the visible wavelength range can be expressed as

$$\text{FoM}_{Vis} = A \times \bar{r}_{Vis} + B \times \bar{V}_{Vis} \tag{4}$$

where $\bar{r}_{Vis}$ represents the average value of the correlation coefficients $r_{ij}$, and $\bar{V}_{Vis}$ indicates the internal differences of the spectral responses $V_i$. $A$ and $B$ are the weighting coefficients that balance the importance of $\bar{r}_{Vis}$ and $\bar{V}_{Vis}$ during the optimization process, specifically set as 10 and 1, respectively. The detailed description of $\bar{r}_{Vis}$ and $\bar{V}_{Vis}$ are shown in Supplement 1, Note S1.

The evaluation functions for near- and mid-infrared (FoM$_{NIR}$ and FoM$_{MIR}$) follow the same formulation as in Equation (4). Spectral responses across the visible, near-, and mid-infrared wavelength ranges are optimized simultaneously, to obtain high encoding efficiency and consequently excellent spectral reconstruction performance across the ultra-broadband wavelength ranges. Therefore, the overall FoM can be expressed as

$$\text{FoM} = \text{FoM}_{Vis} + \text{FoM}_{NIR} + \text{FoM}_{MIR} \tag{5}$$

The detailed description and settings for the PSO method are outlined in Supplement 1, Note S2. Optimization for the varying maximum thicknesses of each individual layer, ranging from 250 nm to 1500 nm, is conducted, generating a variety of film configurations. Fig. 2(a) showcases the evolution of GBest, i.e. the most optimal FoM achieved by the entire swarm, for these varying thicknesses during the optimization procedure. In this case, a lower GBest indicates lower correlation and higher spectral complexity of R$_i$. The PSO method represents an intuitive method for optimizing the film configuration of SSFE from the perspective of spectral response characteristics. Furthermore, it is crucial to assess the efficacy of these optimized film configurations through reconstruction performances, as it offers a more direct and tangible evaluation.

Subsequently, numerical simulations are performed, aiming to find the optimal balance between minimizing the total film thickness, i.e. reducing deposition time and manufacturing costs, while simultaneously achieving prominent performance across ultra-broadband wavelengths. The comprehensive assessment involves reconstructing broadband, single-/dual-peak narrowband spectra. For broadband spectral reconstruction in the visible wavelength range, simulations are conducted using the hyperspectral image (HSI) datasets CAVE[44] and ICVL[45]. The number of spectral channels in CAVE and ICVL is expanded to 301 through linear interpolation within 400 nm to 700 nm wavelength range with 1 nm spacing. To evaluate the reconstruction accuracy, mean square error (MSE), peak signal-to-noise ratio (PSNR), and average structural similarity (SSIM) for all spectral channels are set as key evaluation metrics. Detailed descriptions of these metrics are presented in Supplement 1, Note S3. The error maps showcasing the MSE of each pixel in the HSIs of the upper limit of 1000 nm are shown in Fig. 2(b), with results for other thickness upper limits provided in Supplement 1, Fig. S3. With the thickness of each layer limited to 1000 nm, the PSNR exceeds 35dB and the SSIM consistently surpasses 0.97, indicating a high level of fidelity in the spectral reconstruction of broadband spectra. In Fig. 2(c), we display the reconstructed and the reference spectra of the RGB patches in Fig. 2(b). Remarkably, the average MSE is maintained within the $10^{-5}$ magnitude, demonstrating exceptional reconstruction accuracy. Additionally, single-/dual-peak narrowband spectra are also reconstructed. The variations in the average MSE with different thickness upper limits are depicted in Fig. 2(d-e). The gray planes within these figures indicate levels at which accurate spectral reconstruction is achieved, with values of $5.0\times10^{-4}$/$5.0\times10^{-5}$ for single-/dual-peak narrowband spectra reconstruction, respectively. It is observed that the average MSE generally decreases as the maximum thickness increases, demonstrating an overall trend towards improved spectral accuracy with thicker film layers. However, as the maximum thickness increases, the reconstruction accuracy may deteriorate, as shown in Fig. 2(e). In such circumstance, highly complex spectral responses are regarded as white noise, which disrupts the clear correlation between $I_i$ and the unknown spectra during the spectral encoding process. Considering the results above, it is noted that a maximum thickness of 1000 nm is already sufficient to achieve precise reconstruction of single-peak spectra with a linewidth of 0.5 nm, as well as dual-peak spectra with peaks spaced 3 nm apart, as shown in Fig. 2(f-g) and Supplement 1, Fig. S4, consistent with that observed at greater maximum thicknesses. Numerical tests for near- and mid-infrared wavelength ranges with a maximum thickness of 1000 nm are detailed in Supplement 1, Fig. S5 and Fig. S6 as well. Based on the aforementioned comprehensive performance evaluation and comparison, an optimized encoder configuration with a maximum thickness limit of 1000 nm is selected.

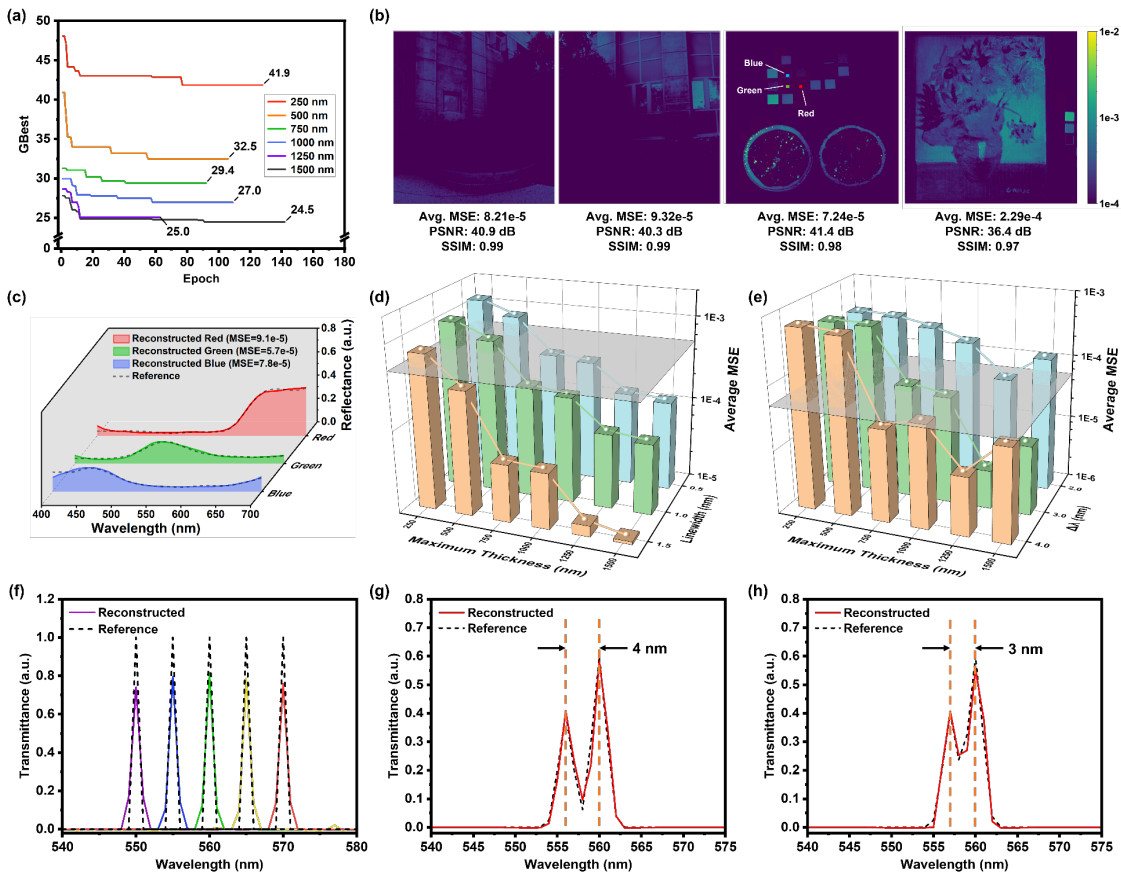

**Fig. 2.** Design of SSFE with the PSO method. (a) Variations of GBest during the iteration process with different maximum thicknesses of each layer within SSFE. The optimization automatically stops when GBest remains unchanged for 50 consecutive iterations. (b) The error maps for reconstructing the HSIs

in CAVE and ICVL using the optimized film configuration with an upper thickness limit of 1000 nm. Evaluation metrics, e.g. MSE, PSNR and SSIM, are calculated for each HIS. (c) The reconstructed and reference spectra of the RGB patches in Fig. 2(b), the MSEs are $9.1\times10^{-5}$, $5.7\times10^{-5}$, and $7.8\times10^{-5}$, respectively. (d) Relationship between the average MSEs and maximum thickness limits of each individual layer for reconstructing the single-peak narrowband spectra with 1.5 nm, 1 nm, and 0.5 nm linewidths. The gray plane indicating accurate spectral reconstruction is set as $5.0\times10^{-4}$. (e) Relationship between the average MSEs and maximum thickness limits of each individual layer for reconstructing the dual-peak narrowband spectra with peaks spaced 4 nm, 3 nm, and 2nm apart. The gray plane indicating accurate spectral reconstruction is set as $5.0\times10^{-5}$. (f) Reconstruction of single-peak narrowband spectra with 0.5 nm linewidth. The average MSE is $2.9\times10^{-4}$. (g-h) Reconstruction of dual-peak narrowband spectra with peaks spaced 4nm and 3 nm apart. The average MSEs are $4.7\times10^{-5}$ and $4.7\times10^{-5}$, respectively.

## 2.3 Fabrication and characterization of the single-spinning film encoder

Based on the preceding analysis, the thickness limit of each individual layer during the optimization procedure is set as 1000 nm. The optimized thickness distribution for the 10-layer SSFE is depicted in Fig. 3(a), with a total thickness of 7393 nm. Particularly, the thicknesses for the $TiO_2$ and $SiO_2$ layers are 4097 nm and 3296 nm, respectively. The encoder is fabricated with one-step electron beam evaporation (EBE) process, specifically selected to bypass micro- and nano-fabrication, thereby reducing the manufacturing complexity and cost. The deposition rates for $TiO_2$ and $SiO_2$ are 0.5 nm/s and 0.8 nm/s, respectively, and the entire deposition process totaling about 3.5 hours. Due to the exclusive application of the mature thin film deposition technique, the optical performance of SSFE is highly consistent and repeatable, eliminating the need for retesting the spectral responses and retraining of the reconstruction network. Following this, angle-dependent spectral responses are measured under different polarization states using both a spectrophotometer (for visible and near-infrared wavelengths) and FTIR (for mid-infrared wavelengths), shown in Fig. 3(b) and Supplement 1, Fig. S7. On the 4-inch-diameter SSFE, the uniformity is evaluated by measuring the transmittance across several distinct positions, shown in Fig. 3(c). Extremely low correlation coefficients of the spectral responses in the visible wavelength range can be obtained with an average value of 0.19, as shown in Fig. 3(d). By simply spinning SSFE, the spectral responses at adjacent angles, e.g. $R_{p,0°}$ and $R_{p,10°}$ (the correlation coefficient is 0.95), or $R_{p,40°}$ and $R_{p,45°}$ (the correlation coefficient is 0.53) exhibit relatively high similarity. In addition, the correlation between $R_p$ and $R_s$ is more pronounced at smaller angles, e.g. $R_{p,20°}$ and $R_{S,20°}$ (the correlation coefficient is 0.96). Accordingly, measurements are taken at 10° intervals at smaller angles (0°, 10°, and 20°), and reduced to 5° intervals for larger angles (20°, 25°, ..., 70°). The correlation coefficients of the spectral responses in near- and mid-infrared are shown in

Supplement 1, Fig. S8. For the same physical thickness, an increase in wavelength, results in simpler spectral responses and lower correlation, as detailed in Supplement 1, Note S4.

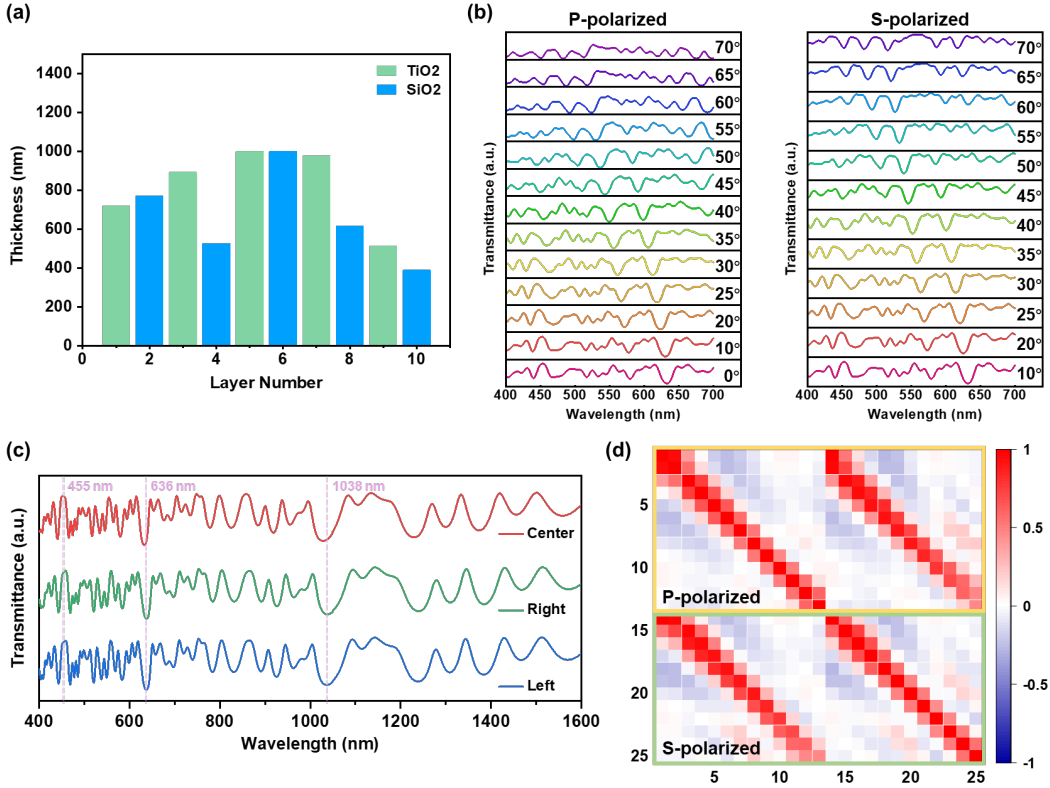

**Fig. 3.** Structural and spectral characteristics of the SSFE. (a) The optimized thickness distribution of the 10-layer SSFE. The first layer is near the substrate side. (b) Measured angle-dependent spectral responses under different polarization states in the visible wavelength range. (c) Measured transmittance across several distinct positions on the 4-inch-diameter SSFE in visible and near-infrared wavelength range. The dashed lines serve as references to highlight the consistency of peak/valley wavelengths. (d) Correlation coefficients of the spectral responses in the visible wavelength range. A remarkably low average correlation of 0.19 is achieved.

## 2.4 Spectral reconstruction in visible, near-, and mid-infrared wavelength ranges

In practical scenarios, system and measurement noises significantly affect reconstruction accuracy. Therefore, we incorporate additional noise to simulate the actual experimental conditions for neural network training, ensuring robustness and improved reconstruction performance. Equation (3) can be rewritten as

$$(\hat{W}, \hat{\theta}) = \arg\min_{W,\theta} \| D(S \cdot R \cdot (1 + \sigma \cdot e)) - S \|_2^2 \tag{6}$$

where $\sigma$ refers to the noise level, which typically varies with different types of detectors and testing environments, and $e$ follows the standard normal distribution. To examine the spectral resolution in the visible wavelength range, we reconstruct the single-peak narrowband spectra with center wavelengths ranging from 500 nm to 600 nm with 10 nm spacing, as shown in Fig. 4(a). The average MSE is $1.05\times10^{-3}$, and the average center-wavelength error and linewidth error are recorded as 0.61 nm and 0.56 nm, respectively, as detailed in Fig. 4(b) and Fig. 4(c). Fig. 4(d-f) displays spectral reconstruction results of the absolute transmittance for color filters. The red/blue solid lines represent the reconstructed spectra using the experimental/calculated intensities. Detailed descriptions of these two intensities are provided in Supplement 1, Note S5. By introducing an ideal reconstructed spectrum, we can find that the reconstruction errors stem from both system inaccuracies (by comparing the reconstructed spectra using the experimental and calculated intensities) and the limited generalizability of the neural network (by comparing the reference spectra and the reconstructed spectra using the calculated intensities). For reconstruction of near-infrared spectra, we numerically reconstruct the transmittance of diverse filters with different spectral distribution, shown in Fig. 3(g-i). Reconstruction performance of broadband spectra is assessed by spectral fidelity, as detailed in Supplement 1, Note S3, demonstrating remarkably high accuracy with an average fidelity of 99.52%. Moreover, in the mid-infrared wavelength range of 3-5 μm, we experimentally reconstruct two broadband spectra generated by a broadband halogen lamp and altering filters, shown in Fig. 3(j-k). The average fidelity drops significantly to 95.61%, much lower than that in the visible wavelength range. The primary reason for the performance decline is the application of the uncooled infrared detector, which generates significant noise and results in a relatively large spectral reconstruction error. To examine the spectral resolution in mid-infrared wavelength range, narrowband spectra with 10 nm linewidths are successfully reconstructed, shown in Fig. 3(l). Herein, the average MSE is $3.36\times10^{-4}$. A comparison of our work with previously reported computational spectrometers is presented in Supplement 1, Table S1. We can observe that our proposed computational spectrometer based on PSEOI principle with SSFE covers an extensive wavelength range from visible to mid-infrared, which is much broader than previously reported works. Additionally, the innovative application of single filter simplifies the overall architecture, potentially reducing manufacturing complexity and costs. The ultra-fine resolution achieved with our spectrometer demonstrates the feasibility of single-filter spectral encoding and the filter design approach based on the PSO method.

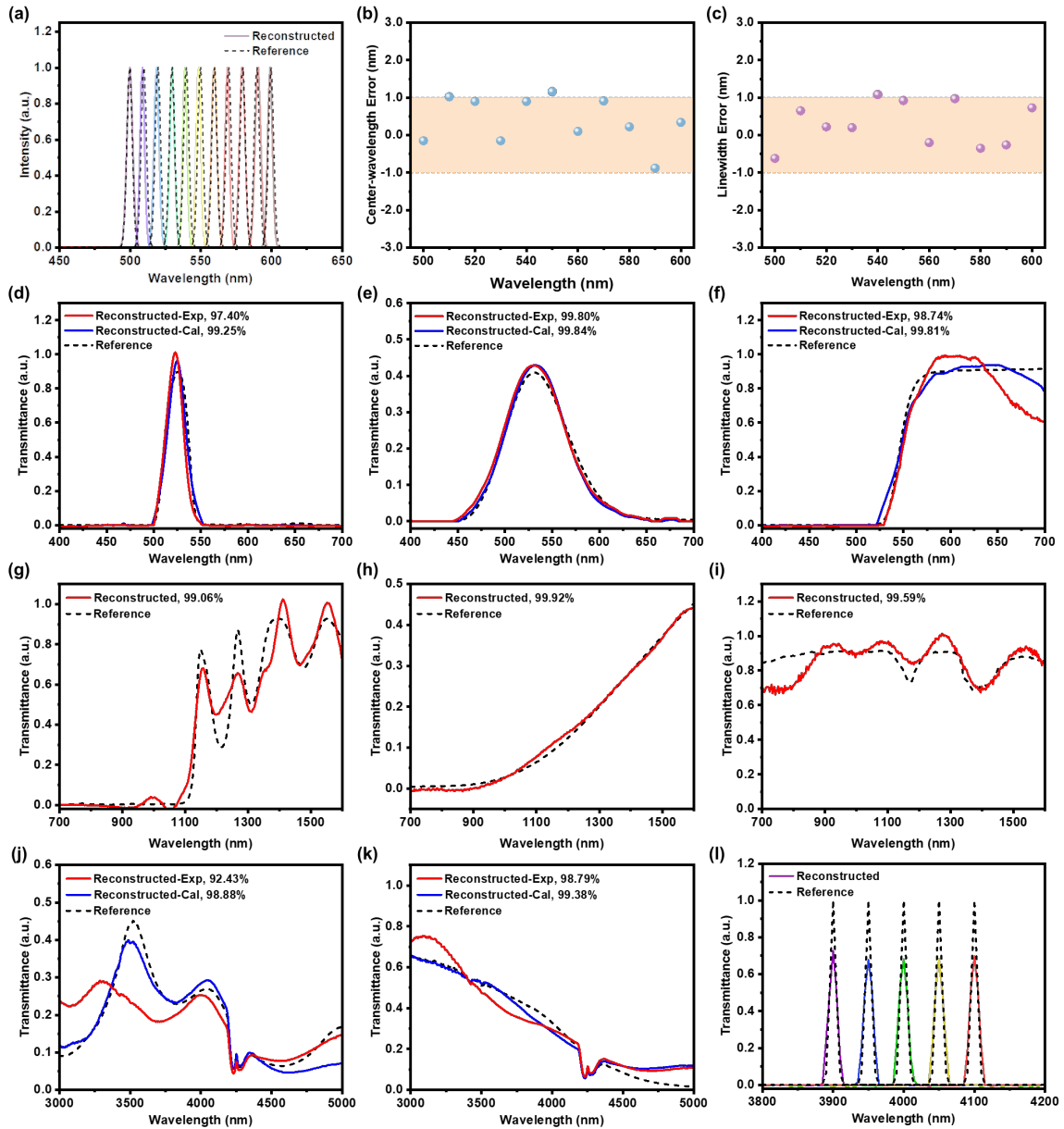

**Fig. 4.** Spectral reconstruction in visible, near-, and mid-infrared wavelength ranges. (a) Reconstruction of narrowband spectra in the visible wavelength range (400 nm to 700 nm with 1 nm spectral channel spacing). The center-wavelength ranges from 500 nm to 600 nm with 10 nm spacing. The black dashed lines represent the reference spectra measured by commercial spectrometer, while the colored solid lines are reconstructed spectra. (b-c) The center-wavelength error and linewidth error distributions for narrowband spectra reconstruction. The average errors are 0.61 nm and 0.56 nm, respectively. (d-f) Reconstruction of broadband spectra. The broadband spectra are the absolute transmittance of color filters. The black dashed lines represent the reference spectra measured by a spectrophotometer, and the red/blue solid lines are the reconstructed spectra using calibrated experimental/calculated intensities.

(g-i) Simulative reconstruction of broadband spectra in the near-infrared wavelength range (700 nm to 1600 nm with 3 nm spectral channel spacing). The black dashed lines represent the reference spectra, while the red solid lines are the reconstructed spectra. (j-k) Reconstruction of broadband spectra in the mid-infrared wavelength range (3 μm to 5 μm with 5 nm spectral channel spacing). (l) Simulative reconstruction of narrowband spectra with 10 nm linewidths.

**2.5 Application in chemical compound classification in mid-infrared wavelength region**

To demonstrate the potential application of the SSFE-based computational spectrometer, we conduct numerical tests to showcase the efficacy of chemical compound classification in the mid-infrared wavelength region. The transmittance of 220 commonly-used liquid chemical compounds in the laboratory constitutes the training/testing dataset (obtained from NIST)[46]. Notably, the dataset is further expanded by introducing spectral variations to simulate different purities and concentrations of the chemical compounds. Fig. 5(a) shows the accurate spectral reconstruction of several chemical compounds with an average fidelity of 99.92%, demonstrating the potential for reliable chemical analysis and identification. The predicted label of the chemical compounds is determined by comparing the reconstructed spectra with the reference spectra in the dataset. For the chemicals in Fig. 5(a), the system correctly identifies the labels, i.e., the types of chemical compounds. However, classification errors do occur, shown in Supplement 1, Fig. S9. It is observed that the fidelities of the misclassifications (with an average fidelity of 99.91%) are not significantly lower than those of correct classifications in Fig. 5(a). The misclassifications primarily stem from the similarity of the spectral profiles among the chemical compounds. Additionally, the transmittance spectra of chemical compounds in mid-infrared often feature notch-shaped characteristics. Compared to the bandpass spectra with peaks at specific wavelengths, spectral reconstruction and classification become more challenging. The classification precision can potentially be enhanced by establishing the straightforward relationship between the labels of the chemical compounds and intensities $I_i$ under distinct polarizations and spinning angles, rather than reconstructing the complete spectra[47,48].

Similarly, we introduce varying levels of noise to the testing dataset to address the measurement and system errors, and to evaluate the robustness of the classification system based on SSFE and deep learning. The confusion matrix of the first 10 chemicals out of the 220 in the dataset is shown in Fig. 5(b). In this case, the noise level $\sigma$ is set as 0.025, and the average precision is 81.38% for 5000 tests. By enhancing the noise level $\sigma$ to 0.05, the average precision drops down to 79.04%, as depicted in Fig. 5(c). Among the 10 compounds, the number of chemicals with identification errors increased from 1 (chemical 6) to 4 (chemicals 5, 6, 8, 10). The trend between the noise levels $\sigma$ and the average precision is illustrated in Supplement 1, Fig. S10.

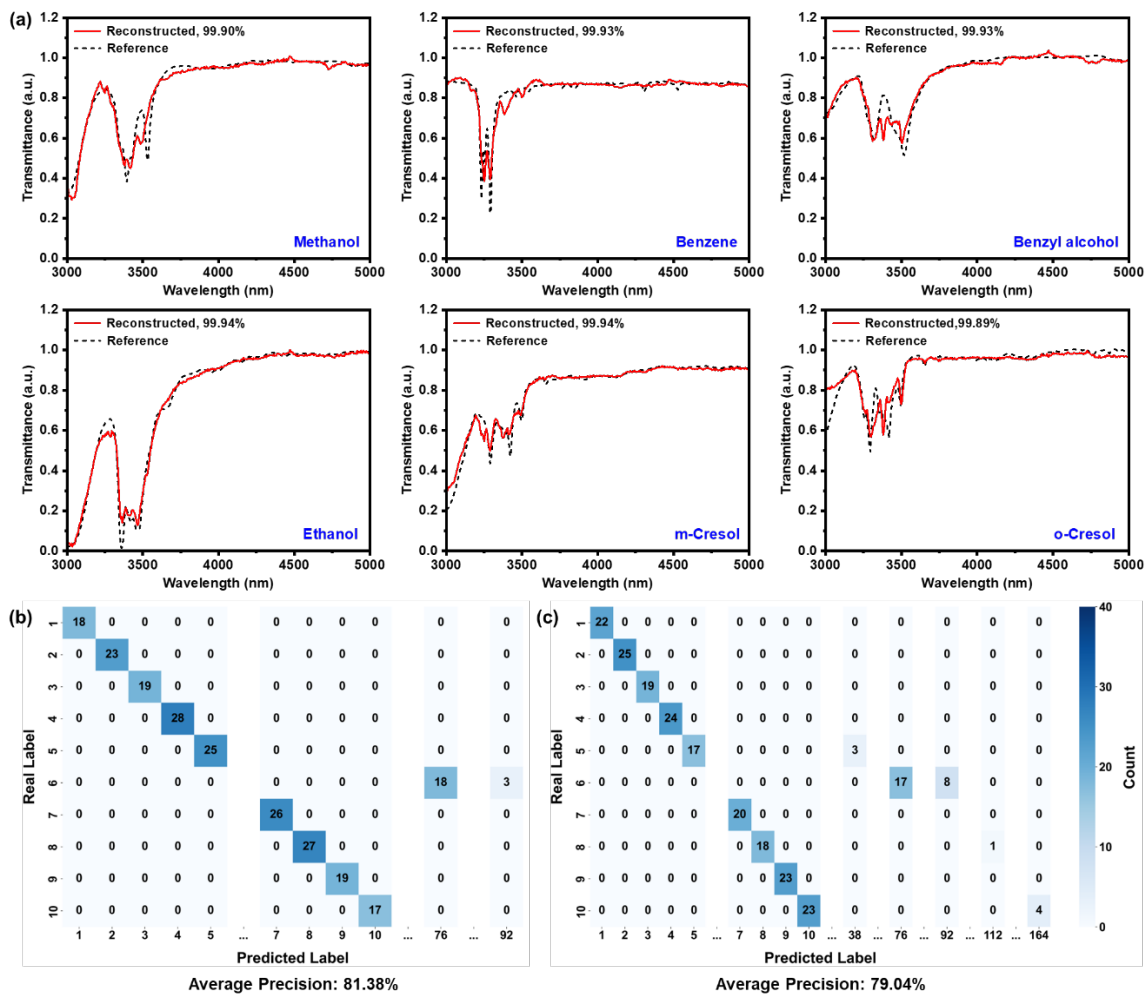

**Fig. 5.** Chemical compound classification in the mid-infrared wavelength region. (a) Reconstruction of the transmittance of chemical compounds. The black dashed lines represent the reference spectra, while the red solid lines are the reconstructed spectra. (b-c) Confusion matrix of the classification results for the first 10 out of the overall 220 chemicals in the dataset with the noise level of (b) 0.025 and (c) 0.05. At the noise level $\sigma$ of 0.025 and 0.05, the average precisions for all chemicals are 81.38% and 79.04%, respectively, across 5000 tests.

## 3. Discussion

In conclusion, we exploit PSEOI for single-filter-based spectral encoding. The implementation of SSFE coupled with a deep learning framework has significantly advanced the field of computational spectroscopy. To the best of our knowledge, it is the first attempt to use a single filter for spectral encoding and one of the few approaches capable of accomplishing computational spectral reconstruction across the visible, near-, and mid-infrared wavelength ranges. The SSFE is carefully predesigned using particle swarm optimization (PSO) method to ensure low correlation coefficient and high complexity spectral responses at different polarizations and spinning angles in the ultra-broadband wavelength range. Extensive spectral reconstruction, encompassing broadband, single-/dual-peak narrowband spectra, is conducted across various spectral bands. The remarkably high fidelities highlight the robust applicability across a variety of spectral characteristics. Furthermore, the potential application to classify chemical compounds, even in the challenging mid-infrared region, underscores the capacity of non-destructive, low-cost, rapid chemical identification and analysis. Nevertheless, spectral imaging application is challenging with the proposed system. As obliquely-incident light passes through the encoder, it undergoes refraction at both the entrance and exit interfaces, resulting in a horizontal shift of the images, shown in Supplement 1, Fig. S11. Therefore, utilizing the intensities $I_i$ from the identical position for each spectral pixel may lead to reconstruction errors, i. e. it is challenging to accurately locate the position of the intensities $I_i$ for each spectral pixel. In future work, it may be possible to incorporate a lens after the encoder to mitigate the shift, thereby exploring the potential for spectral imaging functionality.

## 4. Materials and methods

### 4.1 Characterization of SSFE

The angle-resolved transmittance of the SSFE and the testing samples is measured using the spectrophotometer (Hitachi, UH4150) for the visible and near-infrared wavelength ranges, and the FTIR (Thermo Fisher Scientific, Nicolet iS50R FT-IR) for the mid-infrared wavelength range.

### 4.2 The experimental set up for spectral reconstruction in the visible wavelength range

For spectral reconstruction in the visible wavelength range, a 150-watt Xenon lamp (Microoenerg, CME-SL150) is used as a broadband light source, with the flatter emission spectrum compared to LED lamp. Following the light source, the broadband linear polarizer (Daheng Optics, GCL-050003), the sample under test, the single-spinning encoder, and the monochromatic CMOS image sensor (Hikrobot, MV-CE200-11UM, with Sony IMX183 sensor) are arranged in sequence. The rotation angle is precisely managed by an electronically controlled turntable. Moreover, for single-peak narrowband spectral reconstruction, the monochromatic spectra are generated using the Xenon lamp and a monochromator (Zolix, Omni-λ5005i).

### 4.3 The experimental set up for spectral reconstruction in the mid-infrared wavelength range

For spectral reconstruction in the mid-infrared wavelength range, a halogen lamp (Thorlabs, SLS202L) is used as a broadband light source. Similarly, the broadband linear polarizer (Thorlabs, LPMIR050), the sample under test, the single-spinning encoder, and the InAsSb amplified detector (Thorlabs, PDA07P2) are placed sequentially. The photocurrent signals are read out using a multimeter.

**Declarations**

Data availability

The data and the relevant methods that support this study are available from the corresponding author upon reasonable request.

Conflict of interest

The authors declare that they have no competing interests.

Funding

This work is the supported by Research Funds of Hangzhou Institute for Advanced Study, UCAS (No. 2023HIAS-Y008), and Student Training Program for Innovation and Entrepreneurship of Hangzhou Institute for Advanced Study, UCAS (No. CXCY20230101).

Authors' contributions

The manuscript was written through contributions of all authors. CY and WS conceived the study and supervised the project. JW designed the compact film encoder, JW, YL, WS, CY fabricated and characterized the compact film encoder, JW, WS, CG developed the reconstruction network and performed simulation, JW, WS, CG, SF, YS, HG performed the experiments and analyzed the data. JW and WS wrote the paper, which was then discussed with CY, WS, YS and YZ. All authors approved the final version of the manuscript.

Acknowledgements
Not applicable.

Ethical approval and consent to participate
Not applicable.

Consent for publication
Not applicable.

Corresponding authors
Correspondence to Weidong Shen or Chenying Yang.

Authors' information
Hangzhou Institute for Advanced Study, University of Chinese Academy of Sciences, Hangzhou, 310024, China
Junren Wen, Weiming Shi, Cheng Gao, Shuaibo Feng, Yu Shao, Haiqi Gao, Yuchuan Shao & Chenying Yang

Shanghai Institute of Optics and Fine Mechanics, Chinese Academy of Sciences, Shanghai, 201800, China
Junren Wen, Cheng Gao, Yu Shao, Haiqi Gao & Yuchuan Shao

State key laboratory of Modern Optical Instrumentation, Department of Optical Engineering, Zhejiang University, Hangzhou, 310027, China
Yujie Liu, Yueguang Zhang, Weidong Shen & Chenying Yang

Center of Materials Science and Optoelectronics Engineering, University of Chinese Academy of Sciences, Beijing, 100049, China

Junren Wen, Weiming Shi, Cheng Gao, Shuaibo Feng, Yu Shao & Haiqi Gao

**Supplementary information**

The online version contains supplementary material available at https://doi.org/XXXX.